# BUSCANDO LAS RADIOGALAXIAS MÁS GRANDES

Heinz Andernach [a,*]

[a] Departamento de Astronomía, División de Ciencias Naturales y Exactas,

Universidad de Guanajuato. heinz@ugto.mx

**Resumen**

Se describen los orígenes de la radioastronomía y el descubrimiento de las primeras radiogalaxias, lo que demostró que la emisión en radio de galaxias activas presenta formas muy diversas y puede alcanzar un tamaño muchas veces superior a su extensión óptica. En 1974 se descubrió la primera radiogalaxia "gigante" (GRG en sus siglas en inglés), varias veces mayor que cualquier otra conocida hasta entonces. En 2012, cuando se habían reportado en la literatura alrededor de 100 GRG de este tipo, con tamaños mayores a 1 megapársec (3,3 millones de años luz), el autor empezó su propia búsqueda de GRG y mantiene una lista que cuenta actualmente con casi 7000 GRG, de las cuales más de la mitad fueron descubiertas por él mismo o por sus estudiantes en el Departamento de Astronomía de la Universidad de Guanajuato. Un análisis de los GRG más grandes no revela ninguna propiedad que explique su crecimiento a tales dimensiones. Los recientes avances en radiotelescopios han generado una gran cantidad de imágenes ricas en GRG, pero debido a la complejidad de identificar sus galaxias anfitrionas, solo una parte de estas imágenes puede ser inspeccionada visualmente por humanos. Algoritmos computacionales actualmente disponibles y los proyectos de ciencia ciudadana son propensos a identificaciones erróneas y pasan por alto una fracción sustancial de GRG, por lo que la supervisión de los resultados por expertos es esencial para mantener la fiabilidad de los mismos.

*Palabras clave:* Galaxias; Rastreos en Radioondas; Radiogalaxias.

# SEARCHING FOR THE LARGEST RADIO GALAXIES

**Abstract**

The origins of radio astronomy and the discovery of the first radio galaxies are described which showed that the radio emission of active galaxies is very diverse in shape and can reach a size of many times their optical extent. In 1974 the first "giant" radio galaxy (GRG) was discovered, several times larger than any previously known one. Since 2012, when about 100 such GRGs larger than 1 Megaparsec (3.3 million light years) had been reported in literature, the author is performing his own search for GRGs and maintains a list of currently nearly 7000 GRGs, with more than half of these found on his own or his students at the Departamento de Astronomía of Universidad de Guanajuato. An analysis of the very largest GRGs does not reveal any single property of these that would explain why they could grow to such large sizes. Recent advances in radio telescopes have led to vast amounts of images rich in GRGs, but due to the complexity of identifying their host galaxies only a fraction of these images can be searched with visual inspection by humans. Currently available machine algorithms and citizen science projects are prone to erroneous identifications and also leave unnoticed a substantial fraction of GRGs, such that supervision of the results by experts is essential to produce reliable results.

*Keywords:* Galaxies; Radio Surveys; Radio Galaxies.

## 1. Introduction

Until 1933 the only part of the electromagnetic spectrum used by astronomers was the optical window, covering only half a decade of the more than 14 decades ranging from gamma-rays to the longest radio waves. The radio engineer Karl Jansky (1933), while searching for the origin of a type of noise, called "hiss" at the time, that disturbed transatlantic communications, discovered that our Galaxy, the Milky Way, was a source of low-frequency (20 MHz) radio continuum radiation. Professional astronomers at the time did not show any interest in this discovery, and it was a radio amateur, Grote Reber, who built, completely on his own, the first parabolic radio reflector of 9.6 m diameter with which he was able not only to corroborate Jansky's findings, but also to detect the Sun as a radio emitter. With some difficulty he managed to publish his results in a journal that was recognized by astronomers (Reber, 1944), but it took his own personal efforts to get a few of them interested to pursue the issue.

After the end of World War II the defense radar equipment became used by radio engineers, often without astronomical education, to observe the sky, albeit with a poor angular resolution of several degrees that only allowed to conclude that the radio emission was concentrated along the plane of the Milky Way, but was not sufficient to relate any of the "discrete radio sources" (emission peaks that clearly exceeded the diffuse emission surrounding them) with any optical object. Bolton et al. (1949) succeeded for the first time to locate the position of three such discrete sources to a precision of a few arcminutes (1 arcmin =1/60 of a degree). They achieved this with a "sea interferometer", i.e. a single antenna located on cliffs on the Eastern coast of Australia, and the Western coast of New Zealand, which receives both the direct rays from the source and the ones reflected from the sea while the source is rising or setting. They identified two radio sources with nearby galaxies (Messier 87 and NGC 5128) and the other one with a supernova remnant (the Crab nebula) in our Galaxy.

After the development of radio interferometers, i.e. arrays of several radio antennas spread over a large enough area to obtain angular resolutions close to or better than one arcminute, it turned out that some of the brightest radio sources in the sky (Baade and Minkowski, 1954) originate in faint galaxies at distances which imply their radio luminosity to be much larger than their optical luminosity. Moreover, their radio emission was often found to extend over several times their optical size. In the early 1960s, a few of the brightest radio sources in the sky were found to coincide with starlike objects whose optical spectra could not be interpreted as belonging to any of the known types of stars, until Schmidt (1963) could show that the spectral lines were redshifted by the expansion of the Universe (i.e. of space-time itself) during the light travel time from the object to us. The relative change in wavelength between the observed and emitted spectral line is then called the "redshift", and, together with the rate of expansion of space-time, as quantified by the Hubble constant, this allows one to estimate the distance to the object. These starlike objects were first coined "quasi-stellar objects" (or QSOs), a term that was soon compressed to "quasars". Their redshift would imply even larger distances from Earth than for the previously found "radio galaxies", with enormous optical and radio luminosities (i.e. emitted energy per time) originating in small volumes around their centers. During the following decade this led to the conclusion that for both, radio galaxies and quasars, only the accretion of matter onto a supermassive black hole (with $10^6$ to $10^9$ solar masses, where 1 solar mass = 2 x $10^{30}$ kg) could provide such energies.

Although the term "radio stars" had been used to refer to discrete radio sources, during the 1950s it became clear that none of the bright radio sources coincided with any bright star of our Galaxy, and that the mechanism that causes the radio radiation must be synchrotron radiation produced by relativistic electrons spiraling in a magnetic field due to the Lorentz force. This radiation is characterized not only by a brightness that declines with increasing observing frequency but is also polarized, i.e. the electric and magnetic field vectors of the incoming radiation show a preferred direction as projected on the plane of the sky, such that it can tell us about the orientation of the magnetic field in the source. Synchrotron radiation was first detected optically in 1947 in terrestrial particle accelerators, hence the name of this radiation. In 1956 the optical emission of the Crab nebula supernova remnant and the optical jet of the M87 radio galaxy were indeed found to be polarized, and by 1962 the diffuse radio emission of the Milky Way as well as that of Cygnus A were also found to be polarized, confirming the synchrotron nature of the emission (Gardner and Whiteoak, 1966).

## 2. Radio Galaxies

During the 1970s more versatile radio interferometers became available. These are arrays of various antennas spread over a large area to achieve an angular resolution (but not sensitivity) corresponding to that of an antenna of the total diameter of that area. Examples are the Westerbork Synthesis Radio Telescope (WSRT) in the Netherlands and the Very Large Array (VLA) in New Mexico, USA, which provided images of increasing numbers of radio galaxies, revealing that for most of them the size of their radio emission exceeded by far the optical extent of the host galaxy, and that they tend to have extended emission regions on both sides of the galaxy, the so-called radio lobes, which are "fed" by collimated jets of relativistic particles spiraling in a magnetic field which is of order of milli-Gauss in the jets and of order micro-Gauss in the lobes (1 Gauss = 0.0001 Tesla), i.e. thousands to millions of times weaker than the magnetic field on Earth's surface. This kind of radio morphology was found exclusively in elliptical (or "early-type") galaxies, while in spiral (or "late-type") galaxies the radio emission is essentially distributed over their optical extent and is due to processes related to star formation. The latter are not considered radio galaxies, and only very rare cases of spiral galaxies with radio jets or lobes are known, and no convincing explanation for this fact has been found as yet (Wu et al., 2022).

In what follows, only objects hosted by elliptical galaxies with radio jets and/or lobes will be discussed. While the presence of radio jets is necessary to supply the lobes with emitting particles, radio emission from the jet is actually detected only in a small fraction of radio galaxies and often requires very high sensitivity observations. In much rarer cases (e.g., Messier 87) these jets are detected also at optical or even X-ray wavelengths, also due to the synchrotron process, but due to higher-energy electrons than those responsible for the radio emission.

The radio morphology of radio galaxies can be divided very roughly into two classes, originally introduced by Fanaroff and Riley (1974): in the edge-brightened (FR II) ones the lobes terminate in high radio brightness regions called hotspots, and the edge-darkened (FR I) sources have bright radio jets starting from the host galaxy which gradually fade out into "radio tails" with increasing distance from the host galaxy. The interpretation is that the jets in FR IIs maintain their relativistic speeds well outside the host galaxy, while the jets in FR Is become turbulent and slow down rather close to the host galaxy. The FR I class is also known to have a much larger variety of shapes of their radio emission, many of these likely caused by their presence in galaxy clusters and their motion through the denser intergalactic (or intracluster) medium rich in gas. This causes their jets to be bent by different amounts depending on the so-called ram pressure resulting from this motion, ranging from weak bending in wide-angle tailed (WAT) sources to strong bending in narrow-angle tailed (NAT) sources. The review by Miley (1980) is still instructive to the present day.

## 3. Giant Radio Galaxies

The typical sizes of the radio emission were found to range from about 150 to 300 kpc[1] with only a few examples larger than 500 kpc. We note that all such linear sizes quoted here are assuming the source to lie in the plane of the sky, which is certainly not true, but it gives us the minimum linear size the source must have, since its true orientation in 3D space is not known. For comparison, the distance from the Milky Way to the nearest similar galaxy, the Andromeda Galaxy M31, is 700 kpc, and the typical extent of a rich cluster of galaxies is about 2 Mpc. In a systematic search for larger radio galaxies with the WSRT, Willis et al. (1974) found the first two radio galaxies larger than 1 Mpc, namely 3C 236 and DA 240, with sizes (converted to the currently accepted value of the Hubble constant of 70 km/s/Mpc, the current rate of expansion of the Universe) of 4.3 and 1.5 Mpc. Objects larger than 1 Mpc became known as "Giant Radio Galaxies" (GRGs)[2] and a few years later also the first giant radio quasars (GRQs) were found. Quasars are galaxies with an extremely bright and compact optical nucleus that often outshines its host galaxy such that most quasars appear to us as starlike. GRQs constitute nearly 20 percent of all giant radio sources but in the present paper we shall refer to them as GRGs as well.

3C 236 remained the largest radio galaxy known until 34 years later when Machalski et al. (2008) discovered J1420−0545 with a size of 4.8 Mpc. Their discovery was based on a visual inspection of the publicly available images of the NRAO VLA Sky Survey (NVSS, Condon et al., 1998) where two slightly extended radio sources were seen well separated but parallel to each other, with a fainter radio source midway between them. The latter coincided with a faint galaxy for which the authors obtained an optical spectrum by exposing a total of 2 hours on a 4-m telescope. The resulting redshift of 0.307 and its angular radio size of almost 18 arcmin then implied a linear size of 4.8 Mpc, as projected on the plane of the sky.

The way J1420−0545 was discovered motivated the present author to propose a project to two summer students supervised within a "verano de investigación" (summer internship) in 2012 to inspect all images of the NVSS, plus those of the SUMSS survey (Bock et al., 1999) that covers the rest of the southern sky. Their task was to record the positions of all extended radio sources that looked promising to be large radio galaxies, and then search the NASA/IPAC Extragalactic Database (NED, https://ned.ipac.caltech.edu/) for galaxies near the center of the radio emission that would have a redshift in NED. Within the seven weeks of their work, they were able to double the number of the then known 100 GRGs to about 200 (Andernach et al., 2012) and also provided positions of several thousand of candidate radio galaxies, some of which were later followed up by me with a surprisingly high rate of success in revealing GRGs. Note that to find all GRGs larger than 1 Mpc, and due to the geometry of the Universe, one must scrutinize all radio sources larger than about 2 arcmin. This is because an object of 1 Mpc diameter would first appear smaller with increasing redshift like in a Euclidean space, but in our Universe would reach a minimum angular diameter of 2 arcmin at a redshift near 1.8 and would then again increase in angular size at higher redshifts (where extended radio galaxies are still extremely rare). As a result, of all radio sources larger than 2 arcmin, upon finding their host object and their redshift, only about one-third of these sources turn out to be distant enough to be a GRG. Fortunately, above an angular size of 5 arcmin this fraction is about 60 percent.

---

[1] 1 parsec = 1 pc = $3.09 \times 10^{16}$ m = 3.26 light years

[2] Since about the year 2000 radio galaxies larger than 0.7 Mpc are also considered as "giant" but since there is no physical reason for this threshold, we refer to GRGs as having a minimum size of 1 Mpc here.

## 4. Large Radio Galaxies in Recent Imaging Surveys

Since about 2015 several novel radio interferometers came into operation, like the Low-Frequency Array (LOFAR) centered in the Netherlands and spread over Northern Europe, the Australian Square-Kilometre Array Pathfinder (ASKAP) in Western Australia, and MeerKAT in South Africa, providing images of unprecedented sensitivity and an angular resolution of a few arcseconds (1 arcsec = 1/3600 of a degree) at frequencies below 1 GHz where the diffuse outer lobes of radio galaxies are brighter than at higher frequencies.

Based on one of these surveys, the LOFAR Two-metre Sky Survey (LoTSS), Mostert et al. (2024) reported the discovery of about 2600 GRGs larger than 1 Mpc in the regular survey, and another 130 were found by us in three more deeply observed, but smaller fields, reaching fainter radio galaxies and higher redshifts (Simonte et al., 2024). Mostert et al. used partly machine learning (ML) and partly the contribution of "citizen scientists" to find these GRGs. However, our visual inspection of individual GRGs listed by Mostert et al. (2024) revealed that most of their objects suffer from an overestimate of their radio size, and a small fraction actually has a wrong host galaxy resulting in different redshifts (larger or smaller) compared to their published hosts. This means that the number of true GRGs newly found by Mostert et al. (2024) is about 30 percent smaller than claimed.

The host position can in part be proven using another recent survey, the VLA Sky Survey (VLASS, Lacy et al., 2020) which covers all the sky north of declination -40° at a frequency of 3 GHz and angular resolution of 2.5 arcsec. While in VLASS usually the lobes are undetected, often the only detected source is the radio nucleus, which is due to its radio spectrum being much flatter, i.e. the decrease in brightness with frequency is much weaker than that of the lobes. This radio nucleus may then serve as an indicator of the optical position of the host galaxy. In addition, a systematic inspection of a small portion of the LoTSS DR2 survey region by the present author has shown that up to 30 percent more GRGs can be found in LoTSS images. However, the net time required to first find and then to optically identify them is about 40 hours per 100 $deg^2$, or about 8000 hours for the final LoTSS DR3, to be released in 2025 and to cover almost all the Northern sky with an area of about 20000 $deg^2$.

The ASKAP interferometer in Australia is currently performing a survey called "Evolutionary Map of the Universe" (EMU, Hopkins et al., 2025) which covers the entire Southern sky at 944 MHz with an excellent sensitivity to the faint, diffuse emission from radio galaxy lobes, but it has two times lower angular resolution than that of LoTSS. While this reduces the reliability of optical identifications for small radio galaxies, it is still an ideal survey to discover new GRGs as we have shown in a master thesis directed by the present author (Rodríguez Yáñez, 2024). In this project a region of 260 $deg^2$ that had previously been observed with optical spectroscopy and in X-rays, ASKAP radio images with similar sensitivity to those of the EMU survey were visually inspected to search for extended radio galaxies. A total of over 1000 such radio galaxies were found and optically identified, of which 130 turned out to be GRGs larger than 1 Mpc, which is 26 times the number of previously known GRGs in this area. Apart from that, we could show that over 20 percent of GRG hosts are members of galaxy clusters, and preferably the brightest galaxies at their centers, which contrasts with the paradigm that GRGs can develop only in underdense regions of the Universe.

The recent radio surveys also confirmed the expectation that edge-darkened (FR I) radio galaxies would appear with a larger size when observed with better sensitivity since their outermost tails or lobes were too faint to be detected in previous surveys. However, most of these bent-tailed radio galaxies reside in clusters of galaxies with an intergalactic medium denser than elsewhere in the Universe. Especially for the WAT-type sources, known to be hosted predominantly by brightest cluster galaxies located near the bottom of the gravitational potential of galaxy clusters where the ambient medium is densest, one would not expect to find GRGs among them. Nevertheless, the compilation of large radio galaxies of the present author currently lists 20 such WATs with projected linear sizes between 2 and 3 Mpc, and all except two were found in either LOFAR, EMU, or MeerKAT images. Figure 1 presents an example where EMU reveals fainter lobes further North-East and South-West with a projected linear size of 1.7 Mpc, almost twice as large as in NVSS. The left inset shows the optical host galaxy, and the right inset a higher-resolution 3-GHz VLASS image of the inner 2 arcmin of the source. This shows that their jets can penetrate the intracluster medium and be bent on scales of an entire galaxy cluster, and that many more and larger examples are likely to be found in these, still ongoing surveys.

## 5. Citizen Science, Machine Learning, and Redshifts

The present author was involved in the first citizen science project ("Radio Galaxy Zoo", RGZ) dedicated to the identification of extended radio galaxies, which ran from 2013 through 2019, with its first data release published by Wong et al. (2025). The radio images were drawn from the 1.4-GHz VLA survey "Faint Images of the Radio Sky at Twenty Centimeters" (FIRST, Helfand et al., 2015) with 5.4 arcsec resolution and poor sensitivity for diffuse radio lobes. Shortly after the start of the project it was realized that offering links to lower-resolution surveys like NVSS allowed the volunteers to look for potential large lobes not seen in FIRST. My experience with this project was that a small fraction of volunteers was exceptionally ingenious and meticulous in finding large radio galaxies, but that it was important to verify their results, especially for exceptional sources like GRGs. This required to regularly follow up the findings posted by the volunteers on a web interface that was designed for them to communicate with the scientists. This follow-up led me to the discovery of an unexpectedly large number of new GRGs. However, these results have not yet been incorporated in Wong et al. (2025), and some of these GRGs have meanwhile been found in the LoTSS survey and were reported by other authors.

Both, the LoTSS and EMU survey teams have developed machine learning (ML) algorithms in order to find extended radio galaxies and identify their optical or infrared host galaxies. Such an algorithm was used by Mostert et al. (2024) to discover over 750 GRGs larger than 1 Mpc on LoTSS images.

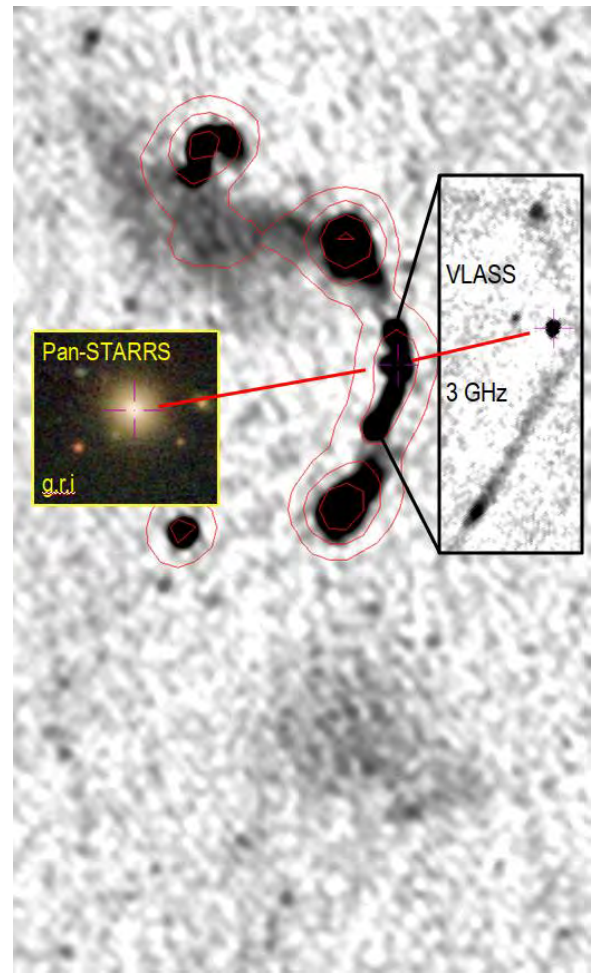

**Figure 1.** 944-MHz image from the EMU survey of the giant WAT J0823+0333 with darker shading indicating stronger radio emission. NVSS contours in red outline its size when first found by citizen scientists (Banfield et al., 2016). North is up and East is left in this and subsequent figures.

While in a majority of cases the result is correct, Figure 2 shows an example of a GRG for which no radio nucleus is obvious in LoTSS, and in which the consideration of higher-frequency and higher-resolution surveys like VLASS would have led to a different choice for the host. In this case Mostert et al. (2024) applied machine learning to propose a faint host galaxy at the tip of the blue arrow in Figure 2, leading to a linear size of 2.0 Mpc for this GRG. However, the only compact radio source detected in VLASS (left inset) coincides with an optical galaxy (right inset) which is also brighter than the ML-based host. With the revised host and its corresponding photometric redshift the size of the GRG is 1.7 Mpc.

The corresponding algorithm for the EMU survey (Gupta et al., 2024) requires a training set that was validated and completed with labels for morphological descriptions from visual inspection by experts. From a comparison of ML results for a small part of the EMU survey with a visual inspection of my own, it became obvious that a large fraction of especially the angularly larger GRGs were missing from the ML results, partly because the ML search was limited to image cutouts of 8 arcmin on a side, and partly because the large lobes of individual GRGs were erroneously recognized as separate radio galaxies and were identified with unrelated galaxies behind these lobes.

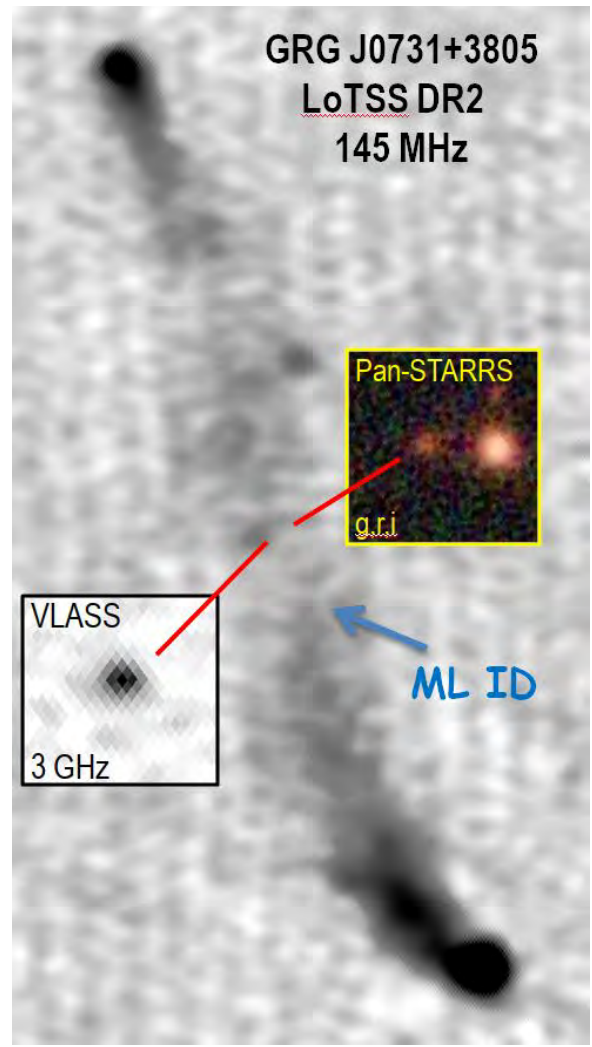

**Figure 2.** 145-MHz image of GRG J0731+3805 in the LoTSS survey. Darker shading indicates stronger radio emission.

An example is shown in Figure 3. To the human eye this is an edge-brightened (FR II) radio galaxy, despite the (rather common) absence of a detected radio jet connecting the host with the lobes. The angular size of the entire source is 6.9 arcmin and its photometric redshift is 0.95, leading to a projected linear size of 3.27 Mpc.

Since GRGs are rare, their host galaxies tend to be faint, implying that for only less than half of these galaxies a spectroscopic redshift is available from which to infer a reliable distance, that would allow one to convert the angular size of the radio emission to their linear size. However, the large-scale spectroscopic surveys performed over the last few decades have led to improved procedures to predict redshifts even for faint galaxies, solely based on the much less time-consuming photometric data from readily available imaging surveys at mid-infrared to ultraviolet wavelengths.

Currently there are well over 100 publications with catalogs of such photometric redshifts (or photo-z's) for a total of 4.6 x $10^9$ galaxies or quasars, albeit with a large overlap between catalogs. Most publications on GRGs only make use of the most readily available such catalogs, while the present author has tried to draw these photo-z's from as many as possible such catalogs. When these were not available from the commonly used databases or catalog browsers in professional astronomy (like VizieR at the CDS Strasbourg, France, https://vizier.cds.unistra.fr/) they were obtained from websites listed in the relevant publications.

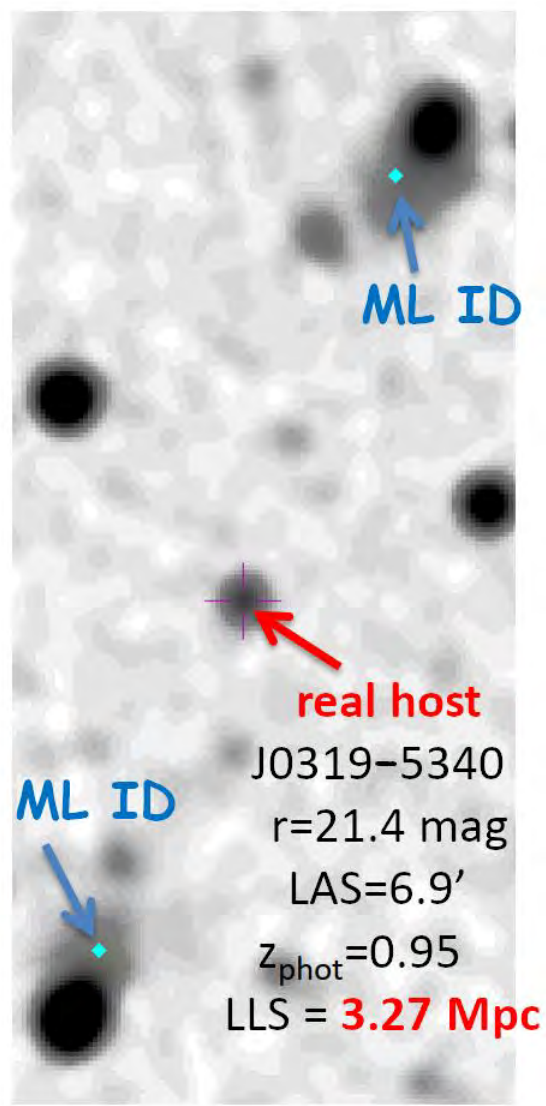

**Figure 3.** 944-MHz EMU image of GRG J0319−5340 with symmetric lobes. The red gapped cross at center indicates the host position. The cyan rhombs on each of the lobes indicate infrared sources which the ML algorithm selected as the hosts for each individual lobe.

Photo-z's can be affected by so-called "catastrophic outliers", i.e. very much over- or underestimated redshifts, e.g. due to lack of sufficient photometric data or failure of the estimation procedure. However, by comparing and averaging several available photo-z's for the same host galaxy their uncertainties can be reduced to under 20 percent in most cases, and outliers can be spotted and excluded. A partial list of such catalogs of photo-z's was provided by Simonte et al. (2024) and Rodríguez Yáñez (2024).

## 6. What allows GRGs to grow that large?

Since 2012 the author has maintained a growing compilation of extended radio galaxies with information on their angular and linear radio size, the position, name, redshift and optical or infrared brightness of their hosts, a rough classification of their radio morphology, and the radio surveys which led to their discovery. The current compilation contains about 30,000 such objects, with a median angular size of 2.0 arcmin, and a median linear size of 0.62 Mpc. These were partly drawn from published literature, but more than half of these were found by the author or his students. The list contains nearly 7000 GRGs with a linear size over 1 Mpc and a median redshift of 0.56. Of these, 790 are larger than 2 Mpc and 150 are larger than 3 Mpc, reaching up to 6.6 Mpc. Only in 2022 the first GRG larger than 5 Mpc had been found, but meanwhile six such GRGs are known, all of them first detected in either LoTSS or EMU images and described in more detail in Andernach and Brüggen (2025). The largest GRG currently known is J0838+5327 with 6.6 Mpc, and the second largest is J1529+6015 with 6.2 Mpc. For the latter Oei et al. (2024) claimed a 3D size of 7 Mpc based on a statistical deprojection. We do not apply this deprojection for the uncertainties involved, and we only quote the projected size on the plane of the sky.

Despite that 51 years have passed since the first discovery of GRGs, there is as yet no clear explanation why these objects can grow to such extraordinary sizes while others do not. We therefore selected the most extreme GRGs from the compilation, namely those larger than a projected size of 3 Mpc, to see whether these could be distinguished by some property from the more modest-sized GRGs. For a sample of 142 such GRGs, 69 of these found by our research group, we used the best available radio survey images to measure (a) their total radio flux and derived their radio luminosity from their redshift, (b) the bending angle between the two opposite lobes, (c) the armlength ratio which we defined as the ratio of the length of the stronger lobe divided by that of the fainter lobe, and (d) the fraction of GRG host galaxies that are members of galaxy clusters, based on cluster catalogs drawn from the most recent deep optical surveys. The results are described in detail in Andernach and Brüggen (2025) and reveal that these extreme GRGs do not differ significantly in any of the above parameters (a) to (d). It is especially striking that the fraction (of about 20 percent) of GRG hosts that are members of clusters of galaxies does not appear to change with their linear size from below 1 Mpc to even above 3 Mpc, despite the fact that one would not expect radio galaxies to grow that large in a denser medium.

## 7. Conclusion

The development of new and more sensitive radio telescopes has provided images of the radio sky that reveal radio galaxies of yet unseen morphology and extent, reaching sizes that exceed those of entire clusters of galaxies. The challenge of these new atlases of the sky lies in the extraction of useful information that can be digested by humans and serves to improve our understanding of the underlying astrophysical processes. Great efforts are being made in terms of computer algorithms (machine learning and other types of supervised or unsupervised learning) as well as citizen science projects in which volunteers are asked to extract the relevant information from a vast amount of images ever increasing in angular resolution and sensitivity. The approach pursued by the present author, namely a systematic visual inspection of small parts of currently available surveys, is clearly impracticable to cope with all available survey images. However, the many examples of erroneous identifications of which only a few are shown here, proves that it remains essential that both machine learning and citizen science require a close supervision by experts to guarantee reliable results. In view of the current tendencies in astronomy towards significant investments in artificial intelligence the formation of human experts must not be neglected.

## Acknowledgements

I thank my colleague Dr. Roger Coziol for comments that improved the manuscript.